\documentclass[12pt]{iopart}
\usepackage{amsfonts}
\begin{document}
\jl{1}
\title[Lorentz invariance and unitarity in UV finite NCQFT]
{Lorentz invariance, Unitarity and UV-finiteness of QFT 
on noncommutative spacetime}
\author{Anais Smailagic\dag{}\footnote[1]{e-mail
 address: \texttt{anais@ictp.trieste.it}},
 Euro Spallucci\ddag{}\footnote[5]{e-mail
 address: \texttt{spallucci@trieste.infn.it}}
 }
 \address{\dag{}Sezione INFN di Trieste,\\
 Strada Costiera 11, 34014 Trieste,\\
 Italy}
\address{\ddag{}Department of Theoretical Physics,\\
University of Trieste, Strada Costiera 11, 34014 Trieste,\\
Italy}
\date{\today}

\begin{abstract}
Ultraviolet finite quantum field theory on even dimensional noncommutative
spacetime is formulated using coordinate coherent states.
$2d$ spacetime is foliated into families of orthogonal,
non commutative, two-planes. Lorentz invariance is recovered if 
one imposes a single non commutative parameter $\theta$ in the theory.
Unitarity is checked at the one-loop level and no violation is found. 
Being UV finite NCQFT does not present any UV/IR mixing. 
\end{abstract}
\maketitle

The interest in noncommutative geometry dates back to \cite{snyder}
where it was pointed out that Lorentz invariance does not exclude the
existence of a \textit{fundamental length} in the spacetime fabric.
In this approach  coordinates are lifted from mere spacetime labels 
to dynamical variables satisfying non-vanishing commutation relations.
In order for some parameter to be given the role of  ``fundamental length''  
it has to be \textit{Lorentz invariant}, i.e. must be inert under Lorentz 
boosts. If so, it can be safely interpreted as a new constant of nature 
representing the ``minimal'' distance having a physical meaning. The existence 
of a universal ``quantum of length'' should provide a natural answer
to many different problems like  curvature singularities in General
Relativity  and ultraviolet (UV) divergences in quantum field theory 
(QFT)\cite{padma}, \cite{pnoi}.\\
Modern string theory re-invented the minimal length through 
introduction of a length invariant under $T$-duality transformations 
\cite{tduality}. It is quite remarkable that non-perturbative string dynamics 
introduces noncommutative (NC) spacetime coordinates as well. 
It seems that there is a deep, yet not completely understood interplay among
Lorentz invariance, noncommutativity (NCY) and the existence of a minimal 
length. Attempts to build up classical, or quantum, dynamics over a NC
space(time) turns out to be a non-trivial task \cite{ag}. 
The most common way out of these difficulties is
to replace NC coordinate operators with standard $c$-number
coordinates and introduce a 
different multiplication rule between ordinary functions known as
Wigner-Weyl-Moyal, or $\ast$-product, \cite{wm}. It has
been originally used in the formulation of non-relativistic quantum 
mechanics (QM) in phase-space. Thus, based on the success in QM, it was
a natural attempt to apply this formulation to quantum field
theory as well. At this point, the problem of compatibility between Lorentz 
invariance and NCY becomes important, but we shall postpone its discussion.
The output of application of the Moyal approach to QFT can be summarized into 
a \textit{deceptively simple} prescription: take a commutative QFT 
and, ``simply'', replace  the ordinary product by 
 the $\ast$-product in the action. The quadratic terms in the action are not
modified by  the $\ast$-product since it adds only surface contributions.
As a result only interaction terms are modified.\\
In spite of its apparent simplicity this prescription opened
a ``Pandora Box'': the $\ast$-product brings into NCQFT
the intrinsic non-locality of the original NC geometry  in the form 
of \textit{non-local} interactions. This non-locality mimics the absence
of ``points'' in the original NC geometry. On the other hand, the presence
of non-local vertices introduces both technical and conceptual problems. 
On the technical side, the
only way to perform calculations of measurable quantities, e.g. 
scattering amplitudes, it to expand the model in powers of the 
``theta-parameter'' measuring the NCY of
the original coordinates. At any finite order in ``theta'' the model
is to all effects an ordinary  \textit{local} field theory (~though with 
additional vertices~), thus completely loosing the memory of its original 
non-locality. The resulting Feynman amplitudes have the same ultraviolet
infinities as in the commutative case. Furthermore, new vertices introduce
 so-called ``non-planar graphs'' leading to an un-expected mixing
between ultraviolet and infrared (IR) divergences. The conclusion is that a
perturbative treatment on the Moyal-starred quantum field theory, not only
is unable to get rid of the UV-infinities, as originally expected, but 
introduces a new kind of problem, i.e. the UV/IR mixing, in conflict with 
renormalization group philosophy. \\
Secondly,  Lorentz invariance requires time to be included among the
NC coordinates, but this may imply problems with unitarity \cite{unit}. 
Several different proposals have been advanced in order to deal with  
the clash between Lorentz invariance and unitarity varying from
attempts to re-define time-ordering \cite{tord}, to \cite{nol} abandoning
Lorentz invariance, to those proposing ``exotic'' treatments 
such as to consider NCY as a sort of ``stochastic'' effect to be averaged
over \cite{carone},\cite{jap1}, \cite{jap2}, \cite{jap3}, \cite{jap4}.
In our opinion, none of them offers neither totally convincing, nor
simple solution.
The skeptical reader can make his opinion  by comparing, 
just to quote some, the conclusions in \cite{demich},\cite{mad1},
 \cite{unit}, \cite{jap3}. 
 \\
Against this background, we proposed in \cite{pi},\cite{pi2} an alternative 
formulation  of NCY in terms of coherent states. Originally,
our primary motivation was to cure short-distance behavior of the Feynman
propagator along the reasoning related to the existence of a minimal length. 
An explicit computation of the vacuum expectation value of the energy-momentum
tensor has shown the infinity suppression mechanism at work \cite{pn}.
 We have shown that the
coherent state approach naturally introduces a Gaussian cut-off in the
Feynman propagator rendering NCQFT UV-finite. Our model was limited to $2+1$
dimensional spacetime, with only space non commutativity. Thus, though it
did produce UV finite theory, it could not lead to 
Lorentz invariant propagator by construction.
In this paper we would like to take on the previous results and improve them
along two important directions outlined in the previous discussion:\\
first, formulate extension of coherent state approach to arbitrary dimensional
space-time. The intent is to discuss physically relevant four dimensional 
NCQFT.\\
Second, we want to obtain a Lorentz invariant and unitary NCQFT.\\

Let us start from the commutator for a set of $D$ {\textit hermitian
operators} $\mathbf{\hat x}^\mu$, $\mu=1\ ,2\ , \dots D $

\begin{equation}
\left[\, \mathbf{\hat x}^\mu\ ,\mathbf{\hat x}^\nu\,\right]= i\,
\mathbf{\hat\theta}^{\mu\nu}
\qquad \mu\ ,\nu= 1\ , 2 \dots\ , D\label{comm}
\end{equation}

Commutator defined in (\ref{comm}) is taken as a starting point in almost all
papers on NCY and all the problems raised depending on how one defines
$\mathbf{\hat\theta}^{\mu\nu}$. Usual choice of $\mathbf{\hat\theta}^{0i}=0$
led to a fixed direction in spacetime and to Lorentz non-invariance.
In order to avoid problems of fixed directions in NC spacetime 
$\mathbf{\hat\theta}^{\mu\nu}$ in (\ref{comm}) must be
chosen as a Lorentz tensor. With such choice it is understood that time
is also a NC coordinate, since it is the only way to hope for a final
Lorentz invariant QFT.
\footnote{We shall work in Euclidean space-time where
Lorentz symmetry is seen as (pseudo) rotations.}\\
The assumption of Lorentz covariance of (\ref{comm}) is crucial. Then, one can 
exploit known property that any antisymmetric matrix can be brought by a 
suitable rotation to a block-diagonal form as 

\begin{eqnarray}
&&\mathbf{\hat\theta}^{\mu\nu}=\mathrm{diag}\left(\,
\mathbf{\hat\theta}_1\ ,\mathbf{\hat\theta}_2\dots \mathbf{\hat\theta}_{D/2}\, 
\right)\\
&& \mathbf{\hat\theta}_i
\equiv \theta_i\, \left(\, \begin{array}{cc}
0 & 1\\
 -1 & 0
\end{array}\,\right)
\end{eqnarray}

In case of odd dimensional space(time) the last element on the diagonal is 
zero, thus in this space(time) there is always one coordinate commuting with all
others. Covariance of (\ref{comm}) allows us to assert that apparent
non-commutativity of all space-time coordinates is actually reducible to a set 
of non-commutative planes.
In other words, NC space-time can always be \textit{foliated} in such a way
that NCY is
restricted to these privileged planes. Lorentz covariance of (\ref{comm}) means
that different inertial observers see different non-commutativity of
coordinates due to different projections of the NC planes in their
respective frames. What they agree on is the existence of these planes in
spacetime.\\
In order to consider full NC we choose to work in even dimensional space
times. In our picture, $D\equiv 2d$ coordinates are represented by $d$ 
two-vectors $\vec \mathbf{\hat x}_i$ as

\begin{eqnarray}
\mathbf{\hat x}^\mu &&= \left(\,
\mathbf{\hat x}^1\ ,\mathbf{\hat x}^2\ , \dots 
\mathbf{\hat x}^{2d-1}\, \mathbf{\hat x}^{2d}\,\right)\nonumber\\
&&=\left(\,
\vec\mathbf{\hat x}_{1}\ , \dots \vec\mathbf{\hat x}_d\,\right)
\end{eqnarray}
where $\vec \mathbf{\hat x}_i\equiv
\left(\,\mathbf{\hat y}_{1\,i}\ ,\mathbf{\hat y}_{2\,i}\,\right)$
are two-vectors with $\left(\,\mathbf{\hat y}_{1\, i}\ ,
\mathbf{\hat y}_{2\,i}\,\right)$
coordinates of the $i$-th NC two-plane satisfying
\begin{equation}
\left[\, \mathbf{\hat y}_{1\, i}\ ,
\mathbf{\hat y}_{2\, i}\,\right]=i\, \theta_i
\end{equation}

Thus, the problem is reduced to an effective two-dimensional one.
Pioneering approach in use of coherent states regarded quantum optics
and referred to phase-space. Coherent states are defined as eigenstates of 
properly defined ladder operators \cite{glauber}. As already stated, the reason
behind use of coherent states is that there
are no coordinate eigenstates for NC coordinates and no coordinate
representation can be defined.
Therefore, ordinary wave functions (in QM) or
fields defined over points (in QFT) can not be defined anymore. 
Coherent states are the closest to the sharp coordinate states that one can
define for NC coordinates in the sense that they are minimal-uncertainty states
and enable, in spite of absence of sharp
coordinate eigenvalues, to define \textit{mean values} of coordinate operators.
\\
In order to apply the coherent state approach we have to build appropriate
 set of mutually commuting \textit{ladder operators built from 
 NC spacetime coordinates only}.
This point is important to stress since often there is a confusion among
readers between coherent states in phase space and our coherent
states in NC coordinate space-time.  Ladder operators should satisfy usual
commutation rules of creation and destruction operators of QM.
Mean values of any operator over coherent states
are {\textit commutative} quantities upon which one can construct the
commutative QFT. In block-diagonal basis of (\ref{comm})
we define ladder operators of the $i$-th plane as

\begin{eqnarray}
a_i&&=
\frac{1}{\sqrt 2}\left(\, \mathbf{\hat y}_{1\, i} +
i \mathbf{\hat y}_{2\,i}\,\right)\\
a_i^\dagger &&=
\frac{1}{\sqrt 2}\left(\, \mathbf{\hat y}_{1\, i}
 - i \mathbf{\hat y}_{2\,i}\,\right)
\end{eqnarray}

The ladder operators satisfy usual canonical commutation rules:

\begin{equation}
\left[\, \mathbf{a}_i\
,\mathbf{a}^{\dagger}{}_j\,
\right]=
\delta_{ij}\, \theta_i
\end{equation}

Normalized coherent states, $\langle \alpha\,\vert\,
\alpha\,\rangle =1$ for the above ladder operators, are defined as

\begin{equation}
\vert\, \alpha\, \rangle\equiv \prod_i
\exp\left[\,\frac{1}{\theta_i }
\left(\, \bar\alpha_i\,
\mathbf{a}_i -\alpha_i\,
\mathbf{a}^{\dagger}_i
\,\right)\,\right]\,\vert\, 0\, \rangle
\end{equation}

where, $\vert\, 0\, \rangle $ is the vacuum state annihilated by
$\mathbf{a}_i$.\\
The basic ingredient of our approach is to associate commutative coordinates
to NC ones as their mean values over coherent states, much in the same
way in which one relates classical and quantum
variables in Quantum Mechanics \cite{messia}. Thus, mean coordinates are

\begin{eqnarray}
&&\langle\, \alpha\, \vert\, \mathbf{\hat y}_{1\, i}\, \vert\, \alpha\,
\rangle={\sqrt 2}\,\mathrm{Re}\,\alpha_i\label{x1}\\
&&\langle\,\alpha\, \vert\, \mathbf{\hat y}_{2\,i}\,
\vert\, \alpha\,
\rangle={\sqrt 2}\, \mathrm{Im}\,\alpha_i\label{x2}
\end{eqnarray}

We associate ordinary functions of mean coordinates 
to operator-valued functions through their mean values as follows: $\langle\,
\alpha\, \vert\, F\left(\, \mathbf{\hat x}\,\right)\, \vert\, \alpha\,
\rangle \equiv F\left(\,\alpha\,\right)$. Now, we can define
a NC version of the Fourier transform as

\begin{equation}
\fl
F\left(\, \alpha\, \right)=\int\prod_{\,i\,=1}^d
\left(\frac{d\vec p_i}
{2\pi}\,\right)
\, f\left(\, \vec p_1\ ,\dots \vec p_d \,\right)\,
\langle\, \alpha\, \vert
\exp\left[\, i \sum_{\,i\,=1}^d
\left(\,{\vec p}\cdot \vec{\mathbf{\hat x}}\,
\right)_i \,\right]\vert\, \alpha\, \rangle
\end{equation}

The novel ingredient of the NC Fourier transform is
the {\textit mean plane wave} : $\langle\, \alpha\, \vert
\exp\left(\, i \,{\vec p}\cdot \vec{\mathbf{\hat x}}\,
\right)\vert\, \alpha\, \rangle $. At this point it is important to stress
the difference between our approach and the usual $\ast$-product. 
In our approach the product among functions of mean coordinates
is not modified, rather  
mean values of operator valued functionals between NC coordinate coherent
states makes NCY visible already at the level of a \textit{single} function
or field. This will turn out to be instrumental in obtaining UV finite
NCQFT, as we shall see later. The explicit form of a NC plane wave
can be calculated, using Hausdorff decomposition, as

\begin{equation}
\fl
\langle\, \alpha\, \vert
\exp\left[\, i \sum_{\,i\,=1}^d\left(\,
{\vec p}\cdot \vec{\mathbf{\hat x}}\,\right)_i
\,\right]
\vert\, \alpha\, \rangle =
\exp\left[ -\sum_{\,i\,=1}^d\left(\, \frac{1}{4}
\theta_i\,
\vec p^{\, 2}_i
+ i\,\left(\, \vec p\cdot\vec{y}\,\right)_i
\,\right)\,\right]
\label{thetawave}
\end{equation}

where, $\vec y_i=\left(\, \Re\,\alpha_{\,i\,}\ , \Im\, \alpha_i\,\right)$ 
is the mean position vector in the $i$-th plane. The NC plane wave
(\ref{thetawave}) has been 
also used in formulating the NC version of the path integral. 
We shall now  calculate the Feynman propagator in the mean value formalism 
using the prescription of \cite{pi}

\begin{eqnarray}
\fl
&&G\left(\, \vec y_1-\vec y^{\,\prime}_1\ ,
 \vec y_2-\vec y^{\,\prime}_2
\dots\ , \vec y_d-\vec y^{\, \prime}_d \ ;
m^2\,\right)\nonumber\\
&&\equiv N\int \left[\, De\,\right]
\prod_{\,i=1}^d
\left[\, Dy_i\,\right]
\prod_{j=1}^d\left[\, Dp_j\,\right]
\times\nonumber\\
&& \exp\left\{\, i\ \sum_{i=1}^d \left(\,
\int_y^{y^\prime}
\vec p\cdot d \vec y\,\right)_i
-\right.
\nonumber\\
&& \int_0^T d\tau \left. \left[\,
e(\tau)\, \left(\, \sum_{\,i\,=1}^d
\vec p^{\, 2}_i +m^2\,\right)
+ \frac{1}{2T}\,\sum_{\,i\,=1}^d
\theta_i
\vec p^{\, 2} _i
\right] \,\right\}
\label{red}
\end{eqnarray}

Direct calculation leads to the final form of the Feynman propagator

\begin{eqnarray}
\fl
&&G\left(\, \vec y_1-\vec y^{\, \prime}_1\ ,
\vec y_2-\vec y^{\, \prime}_2
\dots\ , \vec y_d-\vec y^{\, \prime}_d \ ;
m^2\,\right)\nonumber\\
&&= \int\prod_{\,i\,=1}^d
\left(\, \frac{d\vec p_i}{(2\pi)^2}\,\right)
\, e^{i \sum_{\,i\,=1}^d
\left[\, \vec p \cdot (\, \vec y- {\vec y}^\prime\,)
\right]_i}
G_\theta\left(\, \vec p_1\ ,\dots\ ,
\vec p_d \ ; m^2\,\right)
\end{eqnarray}

where, $G\left(\,\vec p_1\ ,\dots\ ,\vec p_d \ ; m^2\,\right)$
is the momentum space propagator, given by

\begin{equation}
G\left(\,\vec p_1\ ,\dots\ ,\vec p_d \,\right)=
\frac{1}{(2\pi)^{2d}}\frac{1}{\sum_{j=1}^d\vec p^{\, 2}_j +m^2}
\exp\left[\, -\frac{1}{2}\sum_{\,i\,=1}^d \theta_i\,\vec p^{\, 2}_i\,\right]
\label{broken}
\end{equation}

As anticipated above, the use of mean values of coherent states 
leads to the
exponential cut-off in the Green function of NCQFT which makes it UV finite.
At this point it is important to investigate Lorentz invariance of this
result. We see that the denominator is already expressed in terms of Lorentz
invariant length of four momentum. On the other hand, Lorentz invariance of 
the exponential term is not so obvious because parameters $\theta_i$ are 
coupled to two-vectors.
Thus, one should rewrite the exponent, in terms of $\Theta^{\mu\nu}$ tensor, in 
order to discuss invariance properties of the Green function. To clarify this 
point, let us consider the following combination which is a Lorentz scalar: 
$\delta_{\mu\lambda}\mathbf{\hat\theta}^{\mu\nu}\, p_\nu \, 
\mathbf{\hat\theta}^{\lambda\rho}\,
p_\rho $. In the basis where $\mathbf{\hat\theta}^{\mu\nu}$ assumes
block-diagonal form,
$\mathbf{\hat\theta^{\mu\nu}}$ acts as a projector on the planes where the
NCY is non-zero. If we define the projected momentum as $\tilde p^\mu\equiv
\mathbf{\hat\theta}^{\mu\nu} \, p_\nu$ , then

\begin{eqnarray}
\mathbf{\hat\theta}^{\mu\nu}\, p_\nu \, \mathbf{\hat\theta}_{\mu}^{\rho}\,
p_\rho &&= \tilde p^2\nonumber\\
&&=\sum_{\,i\,=1}^d\, \theta_i^2\,\vec p^{\, 2}_i
\label{pproj}
\end{eqnarray}

Comparison between (\ref{pproj}) and (\ref{broken}) shows that
in the latter $\theta_i$ enter in a non-covariant
way. On the other hand, in the same basis the determinant of the matrix
$\mathbf{\hat\theta}^2$ is given by
\begin{equation}
\fl
\mathrm{det}\left[\, \mathbf{\hat\theta}^{\mu\nu}\,
\mathbf{\hat\theta}_{\mu}^{\rho}\,\right]= \prod_{i=1\ ,\dots \ , d}\theta_i^2
\end{equation}

We   define an \textit{effective metric} $\kappa^{\mu\nu}$  as
 
 \begin{equation}
\fl
\left[\, \mathrm{det}\left(\, \mathbf{\hat\theta}^{\mu\nu}\,
\mathbf{\hat\theta}_\mu^\rho\,\right)\,\right]^{-1/4d} \,
\mathbf{\hat\theta}^{\mu\nu}\,
\mathbf{\hat\theta}_{\mu}^{\rho}\equiv \kappa^{\nu\rho}
\label{det}
\end{equation}

$\kappa^{\mu\nu}\, p_\mu\,p_\nu\ne \sum_{\,i\,=1}^d\, \theta_i\,\vec p^{\,
2}_i $ which stands in the exponent of (\ref{broken}). However, if we
assume that the NC parameters $\theta_i$ in different planes
are all \textit{ equal}, i.e.  $\theta_i\equiv \theta $, then

\begin{equation}
\fl
\left[\, \mathrm{det}\, \mathbf{\hat\theta}^{\mu\nu}\,
\mathbf{\hat\theta}_{\mu}^{\rho}\,\right]^{-1/4d} 
\mathbf{\hat\theta}^{\mu\nu}\,
\mathbf{\hat\theta}_{\mu}^{\rho}\,p_\nu\, p_\rho= 
\kappa^{\nu\rho}p_\nu\, p_\rho =\theta\, \delta^{\mu\nu}\, p_\mu \, p_\nu
\end{equation}
and the propagator becomes Lorentz invariant.
The choice of coincident $\theta_i$ gives the form of the effective
metric as 
 \begin{equation}
\kappa^{\mu\nu}= \theta\, \delta^{\mu\nu}
\end{equation}

The same choice  has been implemented as an ansatz in
\cite{mad1}, \cite{putz}. Here, we show how it can be constructed 
using  coherent states.
Thus, Lorentz invariance and UV finiteness of NCQFT is assured 
assuming a unique parameter $\theta$ expressing the spacetime NCY. \\
Regarding UV finiteness,
it is important to remark that the Gaussian cut-off is the intrinsic
ingredient of coherent states NCQFT and \textit{not} an ad hoc
regularization device.   \\
The outcome of the above discussion is  Lorentz invariant Feynman 
propagator 

\begin{equation}
G\left(\, p^2\,\right)=
\frac{1}{(2\pi)^{2d}}
\frac{1}{ p^\mu p_\mu +m^2}
\exp\left[\, -\frac{1}{2}
\theta \, p^\mu\, p_\mu\,\right]\label{gl}
\end{equation}

Now we shall describe one of the important effect of NCY by considering the
equation satisfied by the Green function in (mean)coordinate space. Direct 
calculation, using (\ref{gl}) gives

\begin{equation}
\left[\, -\partial^2 + m^2\,\right]\, G\left(\, x-y\,\right)=
\frac{1}{\left(\, 2\pi\theta\,\right)^2}
\exp\left[ -\frac{\left(\,x-y\,\right)^2}{2\theta}\,\right]
\end{equation}

We see that the effect of NCY in the space of commutative mean coordinates
amounts to the substitution of Dirac delta-function (~point-like source~) by a 
Gaussian function (~ smeared source~). Once again the NCY has produced an 
extension of all point-like structures to smeared ones. It is a natural
consequence of the use of mean values to define spacetime positions.\\ 
Now we address the largely debated, and controversial, question of violation
of unitarity in physical amplitudes of NCQFT. This problem was originally
raised in \cite{unit} and
further discussed in \cite{jap3} with contradictory conclusions. 
Various cures to this problem have been
proposed \cite{wien1}, \cite{liao}, \cite{zeiner}, again with controversial
conclusions.\\
In order to address this problem in our approach, and for the reader to
be able to compare it to previous attempts, we consider a 
two-point amplitude in scalar $\phi^3$ field theory as in \cite{unit}. 
Our calculation  is based on the Feynman propagator (\ref{gl}).
Consider the two-point amplitude given by

\begin{equation}
\fl
\Gamma_4\left(\, p \,\right)=\int\,\frac{d^4q}{(2\pi)^4}
\frac{e^{-\frac{\theta}{2}\,q^2}}{q^2+m^2}
\frac{e^{-\frac{\theta}{2}\,\left(p+q\right)^2}}{\left
(p+q\right)^2+m^2}
\label{gamma4}
\end{equation}

Explicit calculations in QFT rely on a widely adopted method which 
is the Schwinger proper-time
representation of Feynman propagator. Due to the Gaussian cut-off this
prescription has to be suitably modified. We find that the Schwinger 
representation of  (\ref{gl})  is

\begin{equation}
\frac{e^{-\frac{\theta}{2} q^2}}{q^2+m^2}=e^{\theta\, m^2/2}
\int_{\theta/2}^\infty ds
\exp{\left[-s\left(\, q^2+m^2\, \right)\right]}
\label{schwinger}
\end{equation}

 Representation (\ref{schwinger})  is crucial for
 greatly simplifying the actual calculation of any Feynman amplitude. With
 prescription (\ref{schwinger}), amplitude (\ref{gamma4}) becomes

\begin{equation}
\fl
\Gamma_4\left(\, p \,\right)
=e^{\theta\, m^2}\int_{\theta/2}^\infty ds \int_{\theta/2}^\infty dt\int 
\frac{d^4q}{(2\pi)^4}
\exp{\left[-s\left(q^2+m^2\right)-t\left(\left(p+q\right)^2+m^2\right)
\right]}
\end{equation}

Upon integrating internal momentum $q$ (\ref{gamma4}) is recast into

\begin{equation}
\fl
\Gamma_4\left(\, p \,\right)
=\frac{1}{16\,\pi^2}  
e^{\theta\,m^2}\int_{\theta/2}^\infty\, \int_{\theta/2}^\infty\,
\frac{ds\,dt}{\left(s+t\right)^2}exp{\left[-\left(s+t\right)m^2-p^2\frac{s\,
t}{s+t}\right]}
\end{equation}

Introducing a change of variables $s=\left(1-x\right)\lambda$ and
$t=x\,\lambda$ one obtains

\begin{eqnarray}
\Gamma_4\left(\, p \,\right)
=&&\frac{1}{16\pi^2}e^{\theta m^2}\int_0^1 dx\int_{\theta\left(m^2
+x\left(1-x\right)p^2\right)}^\infty
\frac{d\lambda}{\lambda}e^{-\lambda}\nonumber\\
=-&&\frac{1}{16\,\pi^2}e^{\theta\,m^2}\int_0^1 dx\, Ei\left[\,-\theta\left(m^2
+x\left(1-x\right)p^2\right)\,\right]
\end{eqnarray}

where $Ei\left[\, -\theta\,
\left(\, m^2+x\left(1-x\right)p^2\right)\,\right]$ is the
Exponential Integral. One can make use of the explicit expression
for $Ei\left(\, x\right)$ to find the useful form of the amplitude as

\begin{eqnarray}
\fl
\Gamma_4\left(\, p \,\right) =&&
-\frac{1}{16\pi^2}\, e^{\theta\,m^2}\,\left\{\,
\gamma +\int_0^1
dx\ln\left[\, \theta\left(m^2 +x\left(1-x\right)p^2\right)\right]\right.
\nonumber\\
&&\left. +\sum_{n=1}^\infty\frac{(-1)^n}{n\,n!}\int_0^1
dx\, \theta^n\, \left(\, m^2 +x\left(1-x\right)p^2\right)^n\,\right\}
\label{amplitude}
\end{eqnarray}

Now we are ready to investigate the question of unitarity (non)violation
in our framework. As a brief reminder, the unitarity of physical amplitudes 
follows from the probabilistic interpretation of quantum mechanics and is 
encoded into the Optical Theorem

\begin{equation}
2\mathrm{Im}\, \Gamma_{ab} = \sum_n\,\Gamma_{an} \Gamma_{nb}^\dagger
\label{optical}
\end{equation}

where the sum is over all possible intermediate states.\\
Problem with unitarity in  physical amplitude  may arise if the 
above relation is not satisfied.\\
In our case the amplitude in the l.h.s. of (\ref{optical}) is given  by 
(\ref{amplitude}), while the r.h.s. corresponds to the tree-level, on-shell,
 decay of a momentum $p$ scalar particle into two  particles of the same kind.
All calculations, so far, have been done  in Euclidean spacetime. This is our
general prescription to perform virtual momentum integrations in Euclidean
spacetime, where the exponential cutoff has unambiguous sign. Only the final
results are, then, continued into Minkowski spacetime and the physical
implications are discussed. Thus, in order to investigate unitarity  
we Wick rotate (\ref{amplitude}) back to real time. \\ 
The imaginary part in (\ref{amplitude}) arises only from the logarithm due 
to its cut along the real axis. There are two physical regions of momentum
 to be considered, i.e. space-like $(\, p^2 >0\,)$
 and time-like $(\, p^2 <0\,)$ momenta. 
First we consider space-like momenta. In this kinematical region there is no 
imaginary part in  (\ref{amplitude})
since the argument of the logarithm never becomes negative in the 
integration region of $x$. Thus, for space-like momentum (\ref{optical}) is
satisfied since the r.h.s. is also zero due to the energy-momentum conservation.
Unitarity violation in the space-like region does not take place.\\
  To complete the analysis of the unitarity non-violation we shall check the
 relation (\ref{optical}) also in the time-like region by performing an
 explicit calculation of  both r.h.s. and l.h.s. in (\ref{optical}).\\
 For time-like momentum we have $p^2=-\vert\,p^2 \,\vert $, and the 
logarithm has an imaginary part for negative values of its argument
$m^2 -x\left(1-x\right)\vert \, p^2\,\vert $. This is true  for the following
values of the integration variable $x$ :

\begin{eqnarray}
&& x_1 \le x \le  x_2\\
&&x_1=\frac{1-a}{2}\\
&& x_2=\frac{1+a}{2}\\
&&a\equiv \sqrt{1-\frac{4m^2}{\vert \, p^2\,\vert }}
\end{eqnarray}

Thus, for time-like momenta one
finds imaginary part of the l.h.s. of (\ref{amplitude}) to be

\begin{equation}
\mathrm{Im}\Gamma_4\left(\, p \,\right)  =-\frac{1}{16\,\pi}\,e^{\theta\,m^2}
\sqrt{1-\frac{4m^2}{\vert\, p^2\,\vert}}
\label{imgamma}
\end{equation}

which gives the same result as in the commutative case, aside the exponential 
factor keeping track of the NCY. Notice that (\ref{imgamma}) is of particularly 
simple form compared
to the same amplitude computed in various approaches within the $\ast$-product 
formulation. The commutative limit of (\ref{imgamma}) is immediately  obtained
by letting $\theta\to 0$.\\
To complete the check of unitarity we shall calculate the tree level diagram,
i.e. the r.h.s. of (\ref{optical}).
The  way to perform this calculation is to use cutting rules in the
(\ref{amplitude}). Cutting rules correspond to the substitution of internal 
propagators within the Feynman graph by a mass-shell delta-function \footnote{
To avoid misunderstanding, we stress that Dirac delta-functions are still
present in the momentum space since the momenta are commuting from the start.
It is only in position space that Dirac delta-functions are replaced by
Gaussian functions. }
which extracts the imaginary part of the matrix element according to the
prescription:

\begin{equation}
\frac{1}{ q^2 + m^2}\longrightarrow -2\pi i\, \delta\left(\,q^2 + m^2\,\right)
\end{equation} 

Thus, $\Gamma_4(p)$ turns out to be 

\begin{equation}
\Gamma_4^{tree}  = -\frac{4\pi^2}{(2\pi)^4}e^{\theta\,m^2}
   \int\,d^4q \delta\left(p^2+m^2\right)
   \delta\left[\, \left(p+q\right)^2+m^2\,\right]
 \end{equation}
 
 We exploit the fact that
$p^2$ is time-like to choose the rest-frame: $\vec p=0$. Lorentz invariance
of the amplitude, in our approach, guarantees that the result obtained in
the rest-frame will hold in any inertial frame.
 The  calculation is performed by first integrating over $q_0$ component of 
 internal momentum and then over $\vec q$, leading to 
 
 \begin{eqnarray}
 \Gamma_4^{tree}  &&=-\frac{1}{16\pi^2}\, e^{\theta\,m^2}\,
 \int d^3q \, \frac{1}{2\vert\, p_0\, \vert}\delta\left[\,
 {\vec q}^{\, 2} +m^2 -\frac{p_0^2}{4}\,\right]\nonumber\\
  &&=-\frac{1}{8\pi}\,e^{\theta\,m^2}\sqrt{1-\frac{4m^2}{\vert\, p^2\,\vert}}
\label{tree}
\end{eqnarray}

In the (\ref{tree}) we have reconstructed the explicitly Lorentz invariant
form by taking into account  that  in the rest-frame 
$\vert\, p^2\, \vert=p_0^2$. 
Comparison between (\ref{tree}) and (\ref{imgamma}) shows that the unitarity
condition (\ref{optical}) holds also for time-like momentum. 
Our conclusion is that there is \textit{ no unitarity violation} in NCQFT 
based on coherent state approach.
Our statement obviously refers to the model considered and to one-loop
calculations. Further investigation of the question of unitarity non-violation
in more complicated models and at higher loop level are required in order to
confirm this preliminary result on general grounds. Nevertheless, in spite of
its limitations this is the \textit{first} time that the absence of unitarity
problem is proven in a model of NCQFT. We intend to further study this problem
in future papers. 
\\
We give the summary of results on the NCQFT presented in this paper. 
What emerges as a direct consequence of the use of coherent states is
a momentum Gaussian cutoff suppressing UV divergences.
Resulting UV finiteness can be traced back  to the existence 
of ``minimal length'' as an intrinsic property of the NC spacetime
geometry. In coordinate space this property manifests itself in an equally
 simple form: the substitution of coordinate Dirac 
delta-functions with Gaussian functions. This is the reason
why we believe that NCY effects should be always seen, thus even
at the level of free Feynman propagator.\\
We have also shown that   Lorentz invariant results can be obtained
if one assumes  a single parameter $\theta$. In this way,
not only one obtains Lorentz invariant theory, but also reduces to  minimum
the number of free parameters in the model. This is in 
agreement with the  ``philosophy of simplicity'' which requires the minimal
arbitrariness.  Same conclusions, though on the basis of quantum gravity
arguments, regarding a single NC parameter are present in \cite{dfr1}, 
\cite{dfr2}, \cite{dfr3}.\\
Finally, we have investigated  very important issue of unitarity in NCQFT
in view of previous inconclusive discussions. There have been attempts
to attribute the lack of unitarity in NCQFT as their being an ``incomplete''
approximation with respect to string theory. Our philosophy in this paper
is slightly different because NCY in QFT has been introduced much before string
theory, \cite{snyder}, and with different motivations. Therefore, our interest
was to investigate the extension of commutative QFT to its NC version
independently from the origin of NCY itself. We have proven that this extension
can be Lorentz invariant and unitary within the limits of the present model. 
We calculated 
explicitly a one-loop Feynman amplitude for $\phi^3$ scalar NCQFT
and proved that unitarity is preserved.  Contrary to the complicated
analysis in the $\ast$-product approach, our calculation is
particularly simple and transparent thanks to the modification of the
Schwinger proper-time representation of the Feynman propagator
(\ref{schwinger}).  This modification enable us to follow as close as
possible same calculations in the commutative case. 
 The procedure offers no ground to believe that there should be  any 
 departure form one-loop conclusions at higher perturbative order,
 but this point requires further investigation.\\ 

Recently an approach using smeared fields \cite{wien1},\cite{wien2},
\cite{putz}, and based on earlier ideas presented in 
\cite{dfr3}, has led to the same conclusion regarding the absence of
UV/IR mixing and finiteness. \\
Although we agree with those results, there is a slight difference in 
underlying philosophy. The smearing proposed in \cite{dfr3} and followed 
in \cite{putz}, is an ansatz motivated by quantum gravity
arguments, while in our approach it is a direct result of the 
application of coherent state mean values. Furthermore, mentioned authors 
propose to put the smearing only in the interaction terms while keeping the 
kinetic term unchanged. From our point of view, NCY is a geometrical property 
of spacetime and is independent of the particular interactions of matter fields.
To put it ``boldly'', a single car (~free particle~) traveling on the cobbled 
road (~NC spacetime~) feels the bumps without need to crash (~interact~)
into another car. \\

\vfill
\centerline{Acknowledgments}
\vskip 10pt
We would like to thank Prof.T.Padmanabhan for reading the draft of
this note, and for useful discussion on the related issue of zero-point
length.

\end{document}